\documentclass{article}

\usepackage{authblk}
\usepackage{graphicx}
\usepackage{hyperref}

\usepackage{caption}
\usepackage{subcaption}

\author[1]{Marta Girones Sanguesa}
\author[1]{Denis Kutnar}
\author[1]{Bas H.M. van der Velden}
\author[1]{Hugo J. Kuijf}

\affil[1]{Image Sciences Institute, UMC Utrecht}

\title{MixMicrobleed: Multi-stage detection and segmentation of cerebral microbleeds}
\date{5 August 2021}

\begin{document}

\maketitle

\section{Introduction}

Cerebral microbleeds are small, dark, round lesions that can be visualised on T2*-weighted MRI or other sequences sensitive to susceptibility effects \cite{Wardlaw2013,DeGuio2016}. Detection of microbleeds is usually performed visually \cite{Greenberg2009}, with the help of validated visual rating scales such as BOMBS \cite{Cordonnier2009} or MARS \cite{Gregoire2009}. Semi-automated tools to assist with microbleed detection have been developed in the past \cite{Seghier2011,Barnes2011,Kuijf2013b,W2013}. Owing to the blooming effect of microbleeds on MRI, where they appear larger with increasing echo time\cite{Wardlaw2013,McAuley2011}, there have not been many methods focussing on segmentation; since size and volume of microbleeds can change depending on the acquisition settings. Nevertheless, more recent deep learning approaches for microbleed detection, address this as a semantic segmentation task: detection via a method that performs segmentation \cite{Dou2016a,T2021}.

In this work, we propose a multi-stage approach to both microbleed detection and segmentation. First, possible microbleed locations are detected with a Mask R-CNN technique \cite{He2017b}. Second, at each possible microbleed location, a simple U-Net \cite{Ronneberger2015} performs the final segmentation.

\section{Material and methods}

\subsection{Data}
This work used the 72 subjects as training data provided by the ``Where is VALDO?'' challenge of MICCAI 2021 (\url{https://valdo.grand-challenge.org/}). Data consisted of three sequences: T1, T2, and T2*; all aligned in the T2*-space. A binary image including the manual segmentation of microbleeds was provided for every subject.

Data originated from three cohorts and the first number of the subject ID identified the cohort (cohort 1, 2, or 3). The data was split into two separate datasets, according to the slice thickness of the images. Subjects 1** and 3** had T2* images with 3.0~mm slices; and subjects 2** had images with 0.8~mm slices. Two separate models were trained for these dataset splits.

\subsection{Pre-processing}
For every patient, image intensities were normalized using the z-score approach. The data originated from different cohorts and all images were resized to a common field-of-view of 512$\times$512 pixels in-plane.

\subsection{Mask R-CNN}
A pre-trained Mask R-CNN model \cite{He2017b,PyTorch2021} was finetuned to obtain an initial detection and segmentation of the microbleeds. The method uses 2D patches of size 64$\times$64. Because of the small size of the microbleeds, patches were up-sampled with a factor of four to 256$\times$256. This ensured that the microbleeds had a detectable size in the patches. The three different modalities were introduced as three separate channels. Data augmentation with random affine transformations and horizontal flips was used. The model was trained on 80~\% of the data for 15 epochs, with a batch size of 6 and learning rate of 5e-6.

\subsection{U-Net}
A simple U-Net was applied to obtain the final segmentations. A four-channel input was introduced in this case and consisted of the whole slice in the T2* image, including the previous and posterior slice, and the predicted output of the Mask R-CNN. In case of the first and last slice for every image, previous and posterior slices were blank. Data augmentation was defined by random affine transformations and horizontal flips. The model was trained on 80~\% of the data for 50 epochs, with a batch size of 4 and learning rate of 5e-5.

\subsection{Post-processing}
Because microbleeds consist of dark spots in the T2* images, the first threshold was obtained by determining the minimum intensity value of every microbleed present in the dataset, determined in the T2* images. This value was then maximized and applied to the predictions of both models as a filter.

To further reduce the number of false positives, a mask is applied to the U-Net predicted outputs. Visual inspection showed that most false positives occur at the outer boundaries of the brain. The mask consists of cropping the outside of the brain in the T2* image for every patient and applying a dilation to include the borders of the brain.

Finally, the U-Net output is threshold at 0.001 to obtain a binary mask of the microbleeds present in every image.

\subsection{Full prediction pipeline}
\begin{figure}[tbh]
\centering
\includegraphics[width=\textwidth]{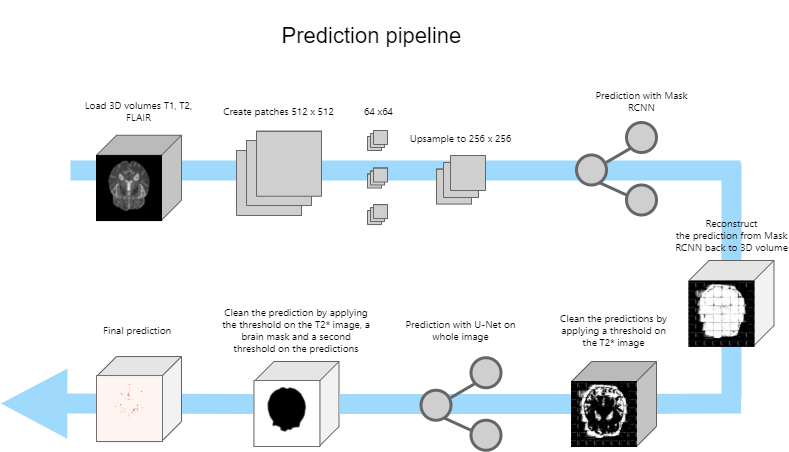}
\caption{Full microbleed prediction pipeline showing the pre-processing, prediction on the models (Mask R-CNN and U-Net) and post-processing steps to detect and segment the microbleeds in a patient.}
\label{fig:MCBS_pipeline}
\end{figure}

To summarize, the final prediction in a subject was obtained by following the next steps: First the image intensity is scaled using z-score normalization and both the images and mask are resized to 512$\times$512. A three-channel input is then introduced in the Mask R-CNN to obtain the first prediction, which is thresholded considering the intensity of the T2* image. This prediction is then included in the U-Net input, together with three consecutive slices in the T2* image, as a four-channel tensor. Finally, the U-Net predictions are threshold with the T2* intensities, masked by eliminating the outside and borders of the brain, and threshold to obtain a binary image. Figure \ref{fig:MCBS_pipeline} shows the full pipeline.

\section{Results}
\begin{figure}[p]
\centering
     \begin{subfigure}[b]{0.45\textwidth}
         \centering
         \includegraphics[width=\textwidth]{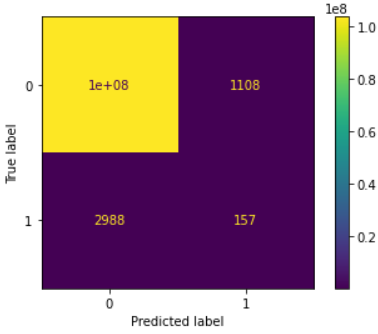}
         \caption{Confusion matrix of subjects 1**.}
         \label{fig:result_1:matrix}
     \end{subfigure}
     \hfill
     \begin{subfigure}[b]{0.45\textwidth}
         \centering
         \includegraphics[width=\textwidth]{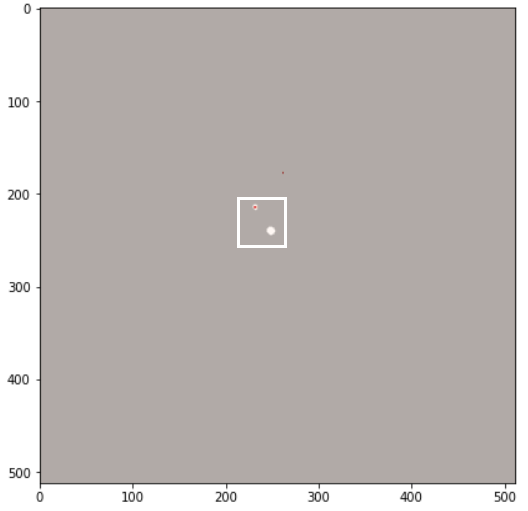}
         \caption{Final segmentation}
         \label{fig:result_1:prediction}
     \end{subfigure}

\hfill

     \begin{subfigure}[b]{0.45\textwidth}
         \centering
         \includegraphics[width=\textwidth]{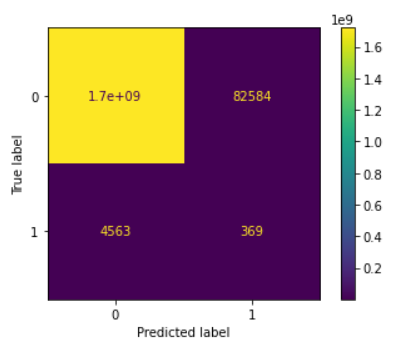}
         \caption{Confusion matrix of subjects 2**.}
         \label{fig:result_2:matrix}
     \end{subfigure}
     \hfill
     \begin{subfigure}[b]{0.45\textwidth}
         \centering
         \includegraphics[width=\textwidth]{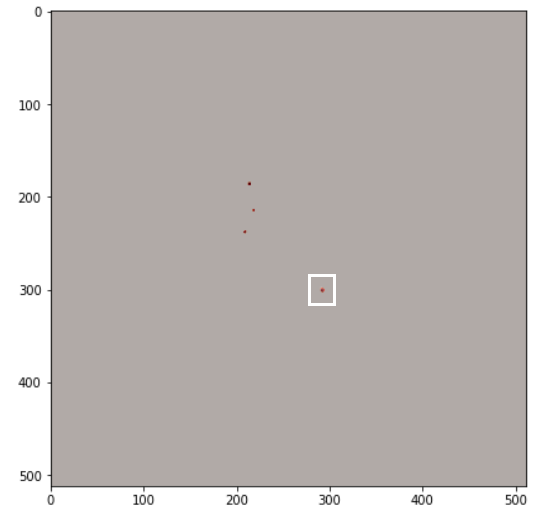}
         \caption{Final segmentation}
         \label{fig:result_2:prediction}
     \end{subfigure}

\hfill

     \begin{subfigure}[b]{0.45\textwidth}
         \centering
         \includegraphics[width=\textwidth]{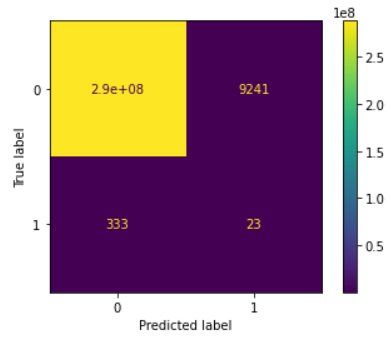}
         \caption{Confusion matrix of subjects 3**.}
         \label{fig:result_3:matrix}
     \end{subfigure}
     \hfill
     \begin{subfigure}[b]{0.45\textwidth}
         \centering
         \includegraphics[width=\textwidth]{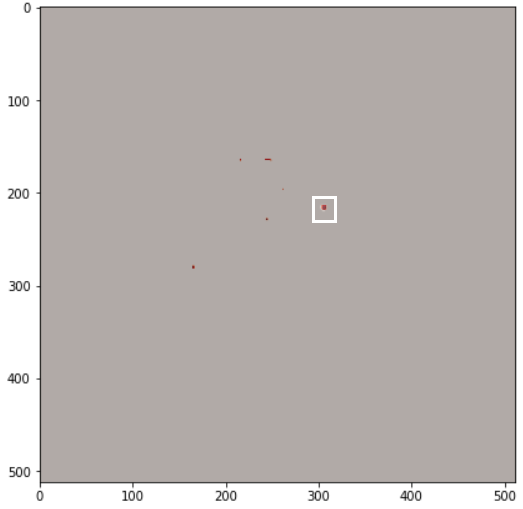}
         \caption{Final segmentation}
         \label{fig:result_3:prediction}
     \end{subfigure}
\caption{Confusion matrices showing the results on the training set of each cohort; and a slice of a random subject showing the true microbleeds (white) and predicted output (red).}
\label{fig:result}
\end{figure}

Figure \ref{fig:result} shows the confusion matrices of subjects 1**, 2**, and 3**, respectively. Subjects 1** and 3** have been processed with the `low' slice thickness model, and subjects 2** with the `high' slice thickness model. 

The prediction obtained from a random subject of each of the cohorts has also been included to visualize the output of the segmentation. Note that, to improve the visualization, the white squares show where the true microbleeds are located in the figure.

\section{Discussion}
\begin{figure}[tbh]
\centering
\includegraphics[width=\textwidth]{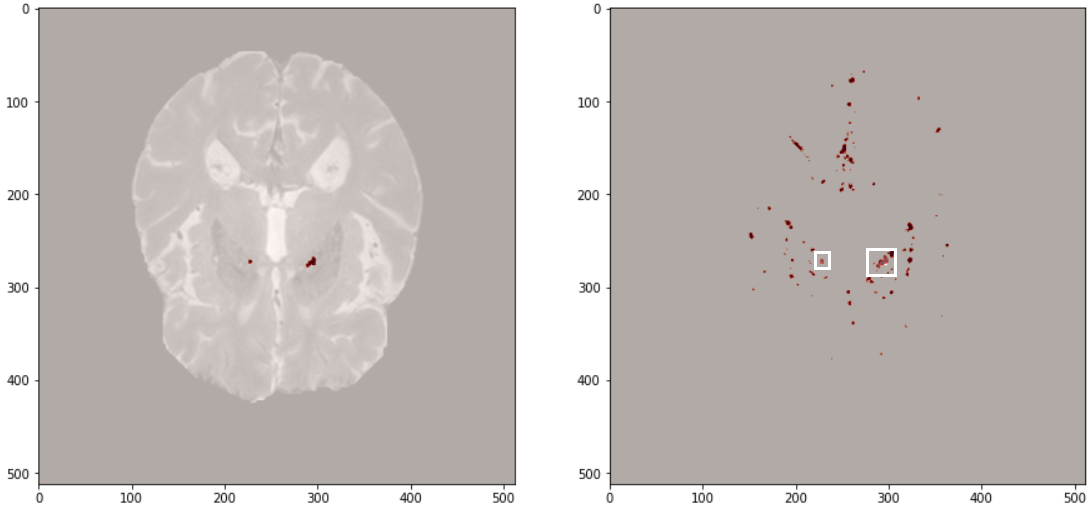}
\caption{Left: slice of a random subject, showing true microbleeds in red. Right: prediction of the method for this slice (red) with the true microbleeds in the white box. A number of false positives are visible, corresponding to dark regions in the T2* image.}
\label{fig:predictions_FP}
\end{figure}

Visual inspection of the results on the training data, revealed that most of the false positive detections are dark areas in the image (Figure \ref{fig:predictions_FP}). This corresponds mostly to locations close to the CSF and vessels present in the brain. The use of masks to remove false positive detections was not enough to clear them all, because of the low intensities (similar to true microbleeds) and the central location. A future implementation could use an improved registration and/or segmentation approach to remove the CSF from the images.

Our current implementation, using a threshold to remove false positive detections, also partially removes the outer boundary of true microbleeds (Figure \ref{fig:predictions_FP}); because the partial volume effect gives it a slightly higher intensity than the signal void at the core of the microbleed. This could be improved by thresholding at the object level (keeping a 3D connected component if at least one voxel survives the intensity threshold) or using a double thresholding and/or region growing approach to retain the borders of true microbleeds.

\begin{figure}[tbh]
\centering
\includegraphics[width=\textwidth]{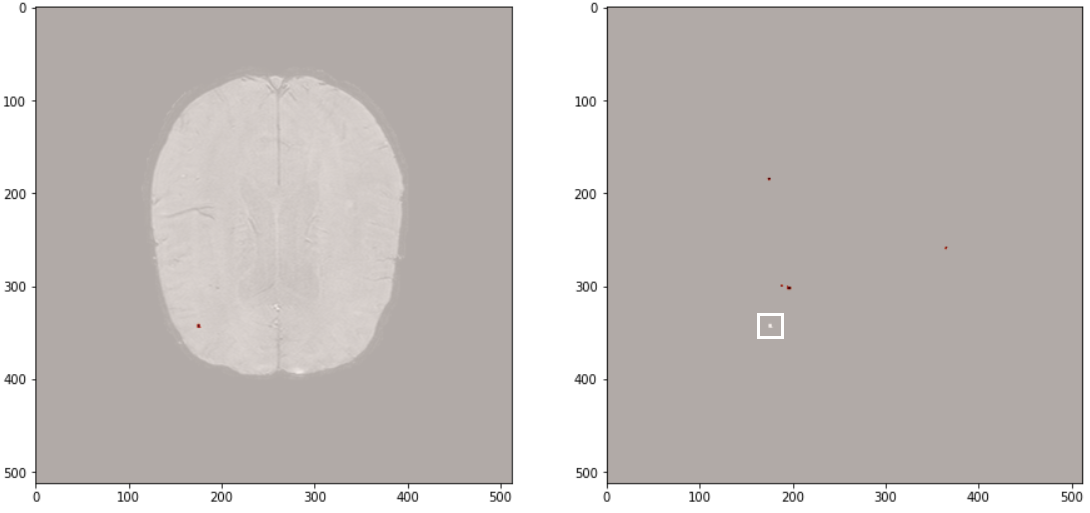}
\caption{Left: slice of a random subject, showing true microbleeds in red. Right: prediction of the method for this slice (red) with the true microbleeds in the white box. The true microbleed was originally detected, but removed by the post-processing steps.}
\label{fig:predictions_FN}
\end{figure}

Similarly, some microbleeds are discard by the last post-processing because they have lower intensities than the threshold, as is shown in the Figure \ref{fig:predictions_FN}.

\section{More information}
Source code is available at: \url{https://github.com/hjkuijf/MixMicrobleed}. The docker container hjkuijf/mixmicrobleed can be pulled from \url{https://hub.docker.com/r/hjkuijf/mixmicrobleed}.

\bibliographystyle{ieeetr} 
\bibliography{}

\end{document}